\documentclass[prb, showpacs]{revtex4}
\usepackage{bm}
\usepackage{graphicx}%
\begin{document}
\title
{Electron-hole pairing in topological insulator heterostructures in
the quantum Hall state}
\author{K.\,V.\,Germash$^1$, D.\,V.\,Fil$^{2}$ }
\affiliation{%
$^1$Karazin Kharkov National University, Svobody Square, 4, Kharkov
61022, Ukraine \\$^2$Institute for Single Crystals, National Academy
of Sciences of Ukraine, Lenin Avenue 60, Kharkov 61001, Ukraine\\ }

\begin{abstract}
A thin film of a topological insulator (TI)  on a dielectric
substrate and a bulk TI - dielectric film -  bulk TI structure are
considered as natural double-well heterostructures suitable for
realizing the counterflow superconductivity.  The effect is
connected with pairing of electrons and holes belonging to different
surfaces of TI and the transition of a gas of electron-hole pairs
into a superfluid state. The case of TI heterostructures subjected
to a strong perpendicular magnetic field is considered. It is shown
that such systems are characterized by two critical temperatures - a
mean-field temperature of pairing and  a much smaller temperature of
the superfluid transition. The dependence of the critical
temperatures on the magnetic field is computed. The advantages of TI
based structures in comparison with GaAs heterostructures as well as
graphene based heterostructures are discussed.
\end{abstract}

\pacs{71.35.Ji; 73.20.-r; 73.22.Gk;}

\maketitle

\section{Introduction}
During the last two decades spontaneous interlayer phase coherence
in quantum Hall bilayers was the subject of comprehensive
investigations. According to the theoretical predictions,
\cite{1,2,3,4,d1,d2} a double-layer electron system subjected to a
strong magnetic field directed perpendicular to the layers should
demonstrate unusual transport behavior at the total filling factor
of Landau levels close to $\nu_T=1$. Such behavior is connected with
the interlayer pairing of electrons and holes belonging to the
zeroth Landau level (the formation of stable magnetoexcitons) and
the transition of the magnetoexciton gas into a superfluid state.
The superfluid state is expected to reveal itself in a flow of
antiparallel electrical supercurrents in adjacent layers and in the
vanishing of the Hall voltage. The effect was realized in AlGaAs
heterostructures by a number of groups.\cite{5,6,7} A huge increase
in the counterflow conductivity and a strong lowering of the Hall
voltage was observed at temperatures below 1 K. Nevertheless, in
these experiments a state with infinite counterflow conductivity was
not registered. It can be accounted for the presence of unbound
vortices,\cite{8,9,10,11}  but the question is still open.

The typical magnetic field used for the observation of the effect in
AlGaAs heterostructures is $B\sim 2$ T. At such a field one can
fulfill the condition for the filling factor $\nu_T=1$ as well as
the requirement for the magnetic length $\ell=\sqrt{\hbar c/e B}$ to
be larger or of order of the interlayer distance $d$. But this field
does not provide smallness of the Coulomb energy
$E_c=e^2/\varepsilon\ell$ ($\varepsilon$ is the dielectric constant
of the matrix) compared to the distance between Landau levels
$\hbar\omega_c=\hbar^2/m_*\ell^2$ ($m_*$ is the effective mass of
carriers). In that case the validity of the lowest Landau level
approximation (commonly used in theoretical studies) is
questionable. The opposite inequality $E_c< \hbar \omega_c$ can be
achieved at higher magnetic fields that corresponds to smaller
$\ell$, but smaller $\ell$ require smaller $d$. Then, the interlayer
tunneling amplitude increases,  that is a negative factor for the
counterflow superconductivity.\cite{12,13,14}

Graphene systems open new prospects for realizing  the
magnetoexciton superfluidity in bilayers.\cite{15,16,17,18,19a,19}
In graphene the relation between the Coulomb energy and the distance
between Landau levels  does not depend on the magnetic field. The
zeroth Landau level in graphene is separated from the nearest
positive and negative levels by the gap $\Delta E_{01}=\sqrt{2}\hbar
v_F/\ell$, where $v_F\approx 10^6$ m/s is the Fermi velocity in
graphene. The inequality $E_c< \Delta E_{01}$ is equivalent to
$\varepsilon> \alpha_{eff}/\sqrt{2}$, where $\alpha_{eff}=e^2/\hbar
v_F\approx 2.2$ is the effective fine structure constant for
suspended graphene. The latter is fulfilled in graphene-based
heterostructures, where SiO$_2$, Al$_2$O$_3$ or BN  compounds are
 used as the dielectric parts.\cite{m1,m2}

As was shown in Refs. \onlinecite{17,18}, an imbalance of filling
factors of graphene layers is required for realizing  the
magnetoexciton superfluidity in bilayer graphene structures. It is
connected with an additional fourfold degeneracy of Landau levels
due to the spin and valley degrees of freedom. This behavior is in
similarity with one for the $\nu_T=2$ quantum Hall bilayers
\cite{nu2}. The imbalance can be created by an electrostatic field
applied perpendicular to the graphene layers.  The change in the
electrical field should follow the change in the magnetic field. It
is required to keep the ratio $E/B$ close to the value $E/B=\alpha
c/\varepsilon$, where $\alpha\approx 1/137$ is the fine structure
constant. For instance, for $\varepsilon=4$ and $B\approx 1$ T the
electrical field $E\approx 5\times 10^3$ V/cm is required.

The discovery of topological insulators (see Refs.
\onlinecite{20,21} and references therein) stimulates new proposals
toward realizing the superfluidity of spatially indirect
excitons\cite{22,23,24,25,26}. The idea\cite{22,23,24,25,26} is that
the surface of TI may work as a natural two-dimensional conductor,
while the interior of TI works as a dielectric. The electron
spectrum of the TI surface states is similar to the graphene
spectrum: it contains Dirac cones. The TI surface should therefore
demonstrate the same quantum Hall behavior as graphene. On the TI
surface the number of Dirac cones is odd, in particular, the most
studied three-dimensional TI Bi$_2$Se$_3$ belongs to a so-called
one-cone family. Thus, TI systems have the same advantage as
graphene systems - the smallness of the Coulomb energy comparing to
the distance between Landau levels, but, at the same time, they are
free from the disadvantage caused by the additional degeneracy of
Landau levels in graphene.

In this paper we analyze two types of structures: a TI film on a
dielectric substrate, and a  bulk TI - dielectric film - bulk TI
heterostructure. In Sec. \ref{s2} we obtain the zero-temperature
phase diagram in the coordinates "the ratio $d/\ell$ - the
dielectric constants." It is shown
 that for the TI film on a substrate  the range of allowed $d/\ell$
  ($d$ is the TI film thickness) is
restricted from below and from above and it shrinks under increase
in the dielectric constant of the substrate $\varepsilon_s$. For
known TIs the dielectric constant $\varepsilon_{TI}$ is large ($\sim
10^2$) and the typical situation corresponds to
$\varepsilon_s/\varepsilon_{TI}\lesssim 0.1$. In this case the state
with electron-hole pairing is expected in the range $d/\ell=0.5 -
2$. For the TI - dielectric film - TI structure the range of allowed
$d/\ell$ ($d$ is the dielectric film thickness) is restricted only
from above and it shrinks under decrease in the dielectric constant
of the dielectric film $\varepsilon_d$. For
$\varepsilon_d/\varepsilon_{TI}\lesssim 0.1$ the maximum allowed
$d/\ell$ is small ($\lesssim 0.2$).

In Sec. \ref{s3} the mean-field critical temperature $T_{mf}$ and
the temperature of the superfluid transition $T_s$ as the functions
of the magnetic field are computed. The temperature $T_{mf}$ is
given by the self-consistence equation for the order parameter. The
temperature $T_s$ is the Berezinskii-Kosterlitz-Thouless transition
temperature. We find that the strong inequality $T_{s}\ll T_{mf}$ is
fulfilled. For the TI film structures the superfluid state is
reached at temperatures higher than for the TI - dielectric film -
TI structures. The temperature $T_s$ is a non-monotonic function of
the magnetic field $B$. At fixed $B$ it increases under decrease in
$\varepsilon_s$ and under increase in $\varepsilon_d$.

\section{Zero-temperature phase diagram}
\label{s2}

Let us consider the electron surface states of a one-cone TI in a
quantizing magnetic field directed perpendicular to the surface. The
low-energy Hamiltonian has the form $H_0=\pm
v_F(p_x\sigma_y-p_y\sigma_x)+m\sigma_z$, where $\sigma_i$ are the
Pauli matrices that act in the spin space, ${\bf
p}=-i\hbar\nabla+e{\bf A}/c$ is the momentum operator, ${\bf A}$ is
the vector potential, $m$ is the Zeeman splitting, and $v_F$ is the
Fermi velocity that is  the material parameter (typically,
$v_F\approx 5\times 10^5$ m/s). The eigenproblem is given by the
Dirac equation
\begin{equation}\label{1}
   \left(
            \begin{array}{cc}
              m & \mp i v_F \left(P_-+\frac{e}{c}A_-\right) \\
              \pm i v_F \left(P_++\frac{e}{c}A_+\right) & -m \\
            \end{array}
          \right)\left(
                   \begin{array}{c}
                     \Psi_\uparrow \\
                     \Psi_\downarrow \\
                   \end{array}
                 \right)=E\left(
                   \begin{array}{c}
                     \Psi_\uparrow \\
                     \Psi_\downarrow \\
                   \end{array}
                 \right),
\end{equation}
where $P_\pm=-i\hbar(\partial_x\pm i\partial_y)$
 and $A_\pm=A_x\pm i A_y$. The upper (lower) sign in Eq. (\ref{1})
corresponds to the top (bottom) surface.

The eigenproblem (\ref{1}) yields the following energies for the
Landau levels: $$E_0=-m, \quad E_{\pm N}=\pm
\sqrt{2\left(\frac{\hbar v_F}{\ell}\right)^2N+m^2},$$ where
$N=1,2,\ldots$. The eigenfunctions are presented in the Appendix.

It is implied that at zero magnetic field the Fermi level is tuned
to the Dirac point $E=0$. It can be done by the appropriate doping
of TI\cite{dop}. In the magnetic field the zeroth Landau level is
shifted from zero. But if the number of surface carriers is
conserved, the Fermi level is shifted as well and it coincides with
the zeroth Landau level. We note that in the case considered the
Zeeman splitting $m=g \mu_B B/2$ (where $\mu_B$ is the Bohr
magneton, and $g$ is the gyromagnetic ratio) is small in comparison
with the distance between the Landau levels (e.g., $m/E_1\approx
3\times 10^{-3}$ at $B=1$ T).

The Coulomb interaction Hamiltonian for the electrons in the zeroth
Landau level reads
\begin{equation}\label{4}
H=\frac{1}{2}\sum_{i,i'}\int d^2 r d^2 r' V_{i,i'}(|{\bf r}-{\bf
r}'|):\hat{\rho}_i({\bf r})\hat{\rho}_{i'}({\bf r}'):,
\end{equation}
where $V_{i,i'}(r)$ is the potential of the Coulomb interaction
 between electrons located on the $i$ and $i'$ working
surfaces,
\begin{equation}\label{5}
    \hat{\rho}_i({\bf r}) =\sum_{X_1,X_2}\quad {\bm \Phi}^*_{0,X_1}({\bf r}) {\bm
    \Phi}_{0,X_2}({\bf r})
    c^+_{i,X_1}c_{i,X_2}
\end{equation}
is the electron density operator in the second quantization
representation, $c_{i,X}^+$ ($c_{i,X}$) is the creation
(annihilation) operator for the electron in the zeroth Landau level
on the surface $i$, $\Phi_{0,X}({\bf r})$ is the eigenfunction [see
the Appendix, Eq. (\ref{2})], and $:\hat{O}:$ means the normal
ordering of an operator $\hat{O}$.

Substituting (\ref{2}) and ({\ref{5}) into (\ref{4}) we obtain the
following expression for the Coulomb interaction Hamiltonian
\begin{equation}\label{6}
    H=\frac{1}{2S}\sum_{ii'}\sum_{X,X',{\bf q}}V_{ii'}(q)e^{-\frac{q^2\ell^2}{2}+iq_x(X'-X)}
    c^+_{i,X+\frac{q_y\ell^2}{2}}c^+_{i',X'-\frac{q_y\ell^2}{2}}
    c_{i',X'+\frac{q_y\ell^2}{2}}c_{i,X-\frac{q_y\ell^2}{2}},
    \end{equation}
where $V_{ii'}(q)$ are the Fourier-components of the  potential and
$S$ is the area of the system.

In what follows we neglect the influence of outer boundaries on the
interaction between  electrons on the working surfaces and consider
the model heterostructure "an infinitely thick dielectric 1 - the
working surface 1 - a dielectric 2 of thickness $d$ - the working
surface 2 - an infinitely thick dielectric 3". In the general case
the dielectrics 1,2, and 3 are characterized by different dielectric
constants $\varepsilon_1$, $\varepsilon_2$, and $\varepsilon_3$,
correspondingly. For such a structure the quantities $V_{ii'}(q)$
read as
\begin{eqnarray}\label{6a}
V_{11}({\bf q})&=&\frac{4 \pi e^2}{q}
\frac{\varepsilon_2+\varepsilon_3+(\varepsilon_2-\varepsilon_3)e^{-2
q d}}{(\varepsilon_2+\varepsilon_3)(\varepsilon_2+
\varepsilon_1)-{(\varepsilon_2-\varepsilon_3)
(\varepsilon_2-\varepsilon_1)} e^{-2 q d}},\\
\label{6b} V_{22}({\bf q})&=&\frac{4 \pi e^2}{q}
\frac{\varepsilon_2+\varepsilon_1+(\varepsilon_2-\varepsilon_1)e^{-2
q d}}{(\varepsilon_2+\varepsilon_3)
(\varepsilon_2+\varepsilon_1)-{(\varepsilon_2-\varepsilon_3)(\varepsilon_2-\varepsilon_1)}
e^{-2 q d}},\\
\label{6c} V_{12}({\bf q})&=&\frac{8 \pi e^2}{q}
\frac{{\varepsilon_2}e^{- q
d}}{{(\varepsilon_2+\varepsilon_3)(\varepsilon_2+\varepsilon_1)}-{(\varepsilon_2-\varepsilon_3)(\varepsilon_2-\varepsilon_1)}
e^{-2 q d}}.
\end{eqnarray}

The pairing of electrons of surface 1 with holes of surface 2 is
characterized by the order parameter
\begin{equation}\label{7}
    \Delta_X=\langle \Psi|c^+_{1X}c_{2X}|\Psi\rangle ,
\end{equation}
where $|\Psi\rangle$ is the many-particle wave function. In
(\ref{7}) the relation between the electron annihilation and hole
creation operator $c_{iX}=h^+_{iX}$ is taken into account. We
consider the many-particle wave function
\begin{equation}\label{8}
    |\Psi\rangle=\prod_X\left(uc^+_{1X}h^+_{2X}+v\right)|{\rm vac}\rangle=\prod_X\left(u c^+_{1X} +v
    c^+_{2X}\right)|0\rangle
\end{equation}
that is an analog of the wave function introduced in the
Bardeen-Cooper-Schrieffer (BCS) theory of superconductivity. The
$u-v$ coefficients satisfy the relation $|u|^2+|v|^2=1$. We
parametrize them as $u=\cos(\theta_0/2)$ and
$v=e^{i\varphi_0}\sin(\theta_0/2)$. In Eq. (\ref{8})
 $|0\rangle$ is
the state with the empty zeroth Landau level, and $|{\rm
vac}\rangle$ is a "vacuum" state defined as $|{\rm
vac}\rangle=\prod_X c^+_{2X}|0\rangle$.

The energy  of the state (\ref{8}) $E=\langle \Psi|H|\Psi\rangle$
reads
\begin{equation}\label{9}
    E=\frac{S}{8\pi\ell^2}\left(W\cos^2 \theta_0
    -\left[\frac{J_{11}+J_{22}}{2}(1+\cos^2\theta_0)+(J_{11}-J_{22})
    \cos\theta_0\right]-J_{12} \sin^2\theta_0
\right),
\end{equation}
where
\begin{equation}\label{10}
    W=\frac{1}{2\pi\ell^2}\lim_{q\to
    0}\left[\frac{V_{11}(q)+V_{22}(q)}{2}-V_{12}(q)\right]=\frac{e^2
d}{\varepsilon_2\ell^2}
\end{equation}
is the energy (per particle) of the direct Coulomb interaction, and
\begin{equation}\label{12}
    J_{ik}=\frac{1}{2\pi}\int_0^\infty q V_{ik}(q)e^{-\frac{q^2
    \ell^2}{2}} d q
\end{equation}
are the energies of the intralayer and interlayer exchange
interaction.

In the state (\ref{8}) the filling factors of the zeroth Landau
level on surfaces 1 and 2 are
\begin{equation}\label{12a}
    \nu_{1(2)}=\frac{1\pm \cos \theta_0}{2} .
\end{equation}
The difference $\tilde{\nu}=\nu_1-\nu_2=\cos \theta_0$ (the filling
factor imbalance) is determined by the condition of minimum of the
energy (\ref{9}) and it may vary under variation in the magnetic
field.

The minimum is reached at
\begin{equation}\label{14}
    \cos \theta_0 =\left\{
                   \begin{array}{ll}
                     1, & \hbox{at $\frac{J_{11}-J_{22}}{2(W+J_{12})-J_{11}-J_{22}}>1$;} \\
                     -1, & \hbox{at $\frac{J_{11}-J_{22}}{2(W+J_{12})-J_{11}-J_{22}}<-1$;} \\
                     \frac{J_{11}-J_{22}}{2(W+J_{12})-J_{11}-J_{22}}, & \hbox{otherwise.}
                   \end{array}
                 \right.
\end{equation}
(it follows from the direct computations that
$2(W+J_{12})-J_{11}-J_{22}>0$ for any $d$ and $\varepsilon_i$).
According to Eq. (\ref{14}) in the general case the filling factor
imbalance depends on the ratio between $\varepsilon_i$ and on the
parameter $d/\ell$.

In the state (\ref{8}) the modulus of the order parameter is equal
to $|\Delta|=\sin\theta_0/2$. Zero imbalance ($\cos \theta_0=0$)
corresponds to the maximum order parameter. This case is realized in
the symmetric heterostructures ($\varepsilon_1=\varepsilon_3$). In
the asymmetric heterostructures the imbalance is nonzero which
results in the lowering of the order parameter. If the imbalance
becomes maximum ($\tilde{\nu}=\pm 1$) the order parameter goes to
zero. The direct evaluation of (\ref{14}) shows that the imbalance
increases under decrease in $d/\ell$ and it reaches the maximum at
some nonzero value of that ratio. Thus, there is the critical
$d/\ell$ below which electron-hole pairing does not occur in the
asymmetric system.

A restriction on the parameter $d/\ell$ also comes  from the
dynamical stability condition (the condition for the collective mode
spectrum to be real valued). To obtain the spectrum of excitations
we follow the approach of Ref. \onlinecite{3au} and consider the
many-particle wave function that accounts for the fluctuations of
the phase and modulus of the order parameter
\begin{equation}\label{15}
    |\Psi\rangle=\prod_X\left(\cos \frac{\theta_X}{2} c^+_{1X} + e^{i \varphi_X}\sin\frac{\theta_X}{2}
    c^+_{2X}\right)|0\rangle .
\end{equation}
In the quadratic approximation the energy of the fluctuations can be
presented in the diagonal form
\begin{equation}\label{16}
    E_{fl}=\frac{1}{2}\sum_q\left(
                   \begin{array}{cc}
                     m_z^*(q) & \varphi^*(q) \\
                   \end{array}
                 \right){\bf K}(q)\left(
                                 \begin{array}{c}
                                    m_z(q) \\
                                   \varphi(q) \\
                                 \end{array}
                               \right),
\end{equation}
where
\begin{equation}\label{17}
    m_z(q)=\frac{1}{2}\sqrt{\frac{2\pi\ell^2}{S}}\sum_X\left(\cos\theta_X-\cos
\theta_0\right)e^{-i q X}
\end{equation}
and
\begin{equation}\label{18}
    \varphi(q)=\sqrt{\frac{2\pi\ell^2}{S}}\sum_X\varphi_X e^{-i q X}
\end{equation}
 are the Fourier-components of the fluctuations of the
filling factor imbalance and the phase of the order parameter,
correspondingly. The matrix ${\bf K}$ in (\ref{16}) is
\begin{equation}\label{19}
    {\bf K}(q)=\left(
                          \begin{array}{cc}
                            K_{zz}(q) & 0 \\
                            0 & K_{\varphi\varphi}(q) \\
                          \end{array}
                        \right)
\end{equation}
with the components
\begin{equation}\label{19a}
 K_{zz}(q)=2\left(H(q)-F_S(q)+F_D(0)+\cot^2\theta_0\Xi(q)\right) ,
\end{equation}
\begin{equation}\label{20}
K_{\varphi\varphi}(q)=\frac{1}{2}\sin^2\theta_0\Xi(q) ,
\end{equation}
where
\begin{equation}\label{21}
    H(q)=\frac{1}{2\pi\ell^2}\left(\frac{V_{11}(q)+V_{22}(q)}{2}-V_{12}(q)\right)
    e^{-\frac{q^2\ell^2}{2}} ,
\end{equation}
\begin{equation}\label{22}
    \Xi(q)=F_D(0)-F_D(q) ,
\end{equation}
\begin{equation}\label{23}
    F_S(q)=\frac{1}{4\pi}\int_0^\infty p
    J_0(pq\ell^2)\left(V_{11}(p)+V_{22}(p)\right) e^{-\frac{p^2 \ell^2}{2}} d
    p ,
\end{equation}
and
\begin{equation}\label{24}
    F_D(q)=\frac{1}{2\pi}\int_0^\infty p
    J_0(pq\ell^2)V_{12}(p) e^{-\frac{p^2 \ell^2}{2}} d p .
\end{equation}
In (\ref{23}),(\ref{24}) $J_0(q)$ is the Bessel function.

The quantities $m_z(q)$ and $\varphi(q)$ are the canonically
conjugated variables. They satisfy the equations of motion
$\hbar\dot{\varphi}(q)=K_{zz}(q)m_z(q)$, $\hbar
\dot{m}_z(q)=-K_{\varphi\varphi}(q)\varphi(q)$ from which the
following expression for the collective mode spectrum comes from
\begin{equation}\label{25}
    \Omega(q)=\sqrt{K_{zz}(q)K_{\varphi\varphi}(q)}.
\end{equation}
Analysis of (\ref{25}) shows that at $d/\ell$ larger than some
critical one the quantity $\Omega(q)$ becomes imaginary valued at
finite $q$. It signals for the dynamical instability of the system.

Thus, for the  TI - dielectric film - TI structure (the symmetric
structure) the range of existence of the superfluid state is given
by the inequality $d/\ell<\tilde{d}_{c1}$, where $\tilde{d}_{c1}$ is
the function of the ratio $\varepsilon_d/\varepsilon_{TI}$. This
function is shown in Fig. \ref{f1}(a). One can see that the range of
$d/\ell$ suitable for realizing the magnetoexciton superfluidity
broaden out under an increase in $\varepsilon_d$. For the TI film on
a dielectric substrate (the asymmetric structure) the range of
existence of the superfluid state is
$\tilde{d}_{c2}<d/\ell<\tilde{d}_{c1}$, where $\tilde{d}_{c1}$ and
$\tilde{d}_{c2}$ are the functions of two ratios:
$\varepsilon_1/\varepsilon_2$ and $\varepsilon_2/\varepsilon_3$.
Here we specify $\varepsilon_1=1$,
$\varepsilon_2=\varepsilon_{TI}=80$ and present the dependencies of
$\tilde{d}_{c1}$ and $\tilde{d}_{c2}$ on the dielectric constant of
the substrate $\varepsilon_s$ [Fig. \ref{f1}(b)]. One can see that
the latter case is in some ways opposite to the previous one:  the
use of dielectric substrates with larger $\varepsilon$ results in
shrinking of the range of allowed $d/\ell$. We emphasize that the
filling factor imbalance can be controlled by the gate voltage
applied to the system. Therefore, in systems subjected to an
electrical field directed perpendicular to the working surfaces the
lower and upper critical $d/\ell$ will differ from ones presented in
Fig. \ref{f1}.

As was already mentioned in the introduction, for graphene-based
heterostructures, in difference with TI heterostructures, the use of
the electrical gate is the necessary condition for realizing the
magnetoexciton superfluidity. The difference is connected with the
presence of only one Dirac cone on the surface of  a TI as compared
to four Dirac cones in graphene. The advantage of the one cone
specifics of TI was discussed previously in connection with
electron-hole paring in the absence of a magnetic field.
\cite{22,23,24,25,26} In such systems the electric gate is in any
case required and the advantage  consists of a reduction of
screening of the interlayer Coulomb attraction between elections and
holes. For the electron-hole pairing in the zeroth Landau level the
screening is not so important. Actually, the effect of screening is
small if the Coulomb energy does not exceed the gap between the
zeroth and $N=1$ Landau level. But  just due to the one cone
specifics of the  TI the state with spontaneous interlayer phase
coherence in the zeroth Landau level is stable with respect to ones
without such coherence. Also, since in difference with graphene and
GaAs heterostructures, the zeroth Landau level in the TI is
completely spin polarized (see the Appendix), low energy excitations
connected with spin (and valley) degrees of freedom are forbidden.
It allows us to consider the TI as a refined system for realizing
the magnetoexciton superfluidity.

\begin{figure}
\begin{center}
\includegraphics[width=8cm]{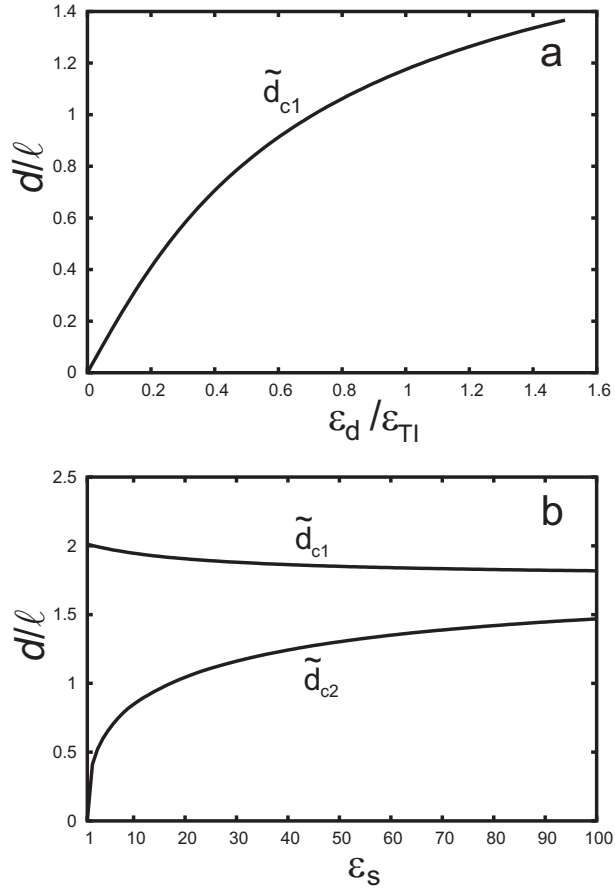}
\end{center}
\caption{The zero-temperature phase diagrams for the TI - dielectric
film - TI heterostructure (a) and for the heterostructure "TI film
on a dielectric substrate"(b).} \label{f1}
\end{figure}

\section{Finite temperature properties}
\label{s3}

 Let us consider finite temperature behavior of the system
in the framework of the mean-field approach \cite{32,33}.

One obtains from (\ref{6}) the following mean-field Hamiltonian
\begin{equation}\label{26}
    H_{MF}=\sum_X\left[\sum_i \epsilon_i c^+_{iX}c_{iX}-(J_{12}\Delta
    c^+_{1X}c_{2X}+\textrm{H.c.})\right],
\end{equation}
where $\Delta=\langle c_{2X}^+ c_{1X}\rangle$ is the mean-field
order parameter,
\begin{equation}\label{27}
  \epsilon_i=D_i-J_{ii}\nu_i-\mu
\end{equation}
 with $\nu_i=\langle c_{iX}^+ c_{iX}\rangle$, the mean-field filling
factors,
 \begin{equation}\label{28}
  D_1=\frac{1}{2\pi\ell^2}\lim_{q\to 0
}\left[V_{11}(q)\left(\nu_1-\frac{1}{2}\right)+V_{12}(q)\left(\nu_2-\frac{1}{2}\right)\right],
 \end{equation}
\begin{equation}\label{29}
D_2=\frac{1}{2\pi\ell^2}\lim_{q\to 0
}\left[V_{22}(q)\left(\nu_2-\frac{1}{2}\right)+V_{12}(q)\left(\nu_1-\frac{1}{2}\right)\right],
\end{equation}
and $\mu$, the chemical potential. In (\ref{28}) and (\ref{29}) the
interaction with the positive background is taken into account.

The Hamiltonian (\ref{26}) is diagonalized using the $u-v$
transformation
\begin{eqnarray}\label{30}
  c_{1X}=u\alpha_X+v^*\beta^+_X,\cr
c_{2X}=u^* \beta^+_X-v\alpha_X,
\end{eqnarray}
where  $u=\cos(\Theta/2)$ and $v=\sin(\Theta/2)e^{i\varphi}$. The
condition of vanishing non-diagonal terms in the transformed
Hamiltonian yields the following relations
\begin{eqnarray}\label{31}
\sin \Theta
=\frac{J_{12}|\Delta|}{\sqrt{\tilde{\epsilon}^2+J_{12}^2|\Delta|^2}},
\quad \cos \Theta
=\frac{\tilde{\epsilon}}{\sqrt{\tilde{\epsilon}^2+J_{12}^2|\Delta|^2}},
\end{eqnarray}
where
\begin{equation}\label{32}
\tilde{\epsilon}=\frac{\epsilon_1-\epsilon_2}{2}=\frac{1}{2}\left[
\left(W-\frac{J_{11}+J_{22}}{2}\right)\tilde{\nu}-\frac{J_{11}-J_{22}}{2}\right].
\end{equation}
The transformed Hamiltonian has the form
\begin{equation}\label{33}
    H_{MF}=\sum_X\left(E_\alpha \alpha^+_{X}\alpha_X+E_\beta
\beta^+_X\beta_X\right)
\end{equation}
with the spectrum
\begin{equation}\label{34}
    E_{\alpha(\beta)}=\sqrt{\tilde{\epsilon}^2+J_{12}^2|\Delta|^2}\pm
\frac{\epsilon_1+\epsilon_2}{2}.
\end{equation}
The condition $\nu_1+\nu_2=1$ yields the relation
\begin{equation}\label{35}
    1=1+N_F(E_\alpha)-N_F(E_\beta),
\end{equation}
where $N_F(E)$ is the Fermi distribution function. The relations
(\ref{34}) and(\ref{35}) lead to the condition
\begin{equation}\label{36}
    \epsilon_1+\epsilon_2=0.
\end{equation}
Note that Eq. (\ref{36}) can be satisfied under the appropriate
choice for the chemical potential $\mu$. It corresponds to that the
chemical potential is determined by the relation $\nu_1+\nu_2=1$.
Under accounting Eq. (\ref{36}) the spectrum (\ref{34}) is reduced
to $E_\alpha=E_\beta={\cal
 E}=\sqrt{\tilde{\epsilon}^2+J_{12}^2|\Delta|^2}$.

Equation (\ref{35}) is the first one in a set of three
self-consistence equations. The other two equations read as
\begin{eqnarray}
\label{37} \tilde{\nu}= -\frac{\tilde{\epsilon}}{{\cal E}}\tanh
\frac{{\cal E}}{2 T},\cr \Delta=\frac{J_{12}\Delta}{2{\cal E}}\tanh
\frac{{\cal E}}{2 T}.
\end{eqnarray}

It follows from (\ref{37}) that at $\Delta\ne 0$ the following
relation is fulfilled
\begin{equation}\label{38}
    J_{12}\tilde{\nu}+2\tilde{\epsilon}=0.
\end{equation}
Equation (\ref{38}) yields
\begin{equation}\label{39}
    \tilde{\nu}=\frac{J_{11}-J_{22}}{2(W+J_{12})-J_{11}-J_{22}}
\end{equation}
that coincides with Eq. (\ref{14}) under assumption that
$|J_{11}-J_{22}|<2(W+J_{12})-J_{11}-J_{22}$. If the latter
inequality is not fulfilled, Eq (\ref{38}) cannot be fulfilled as
well, and the order parameter $\Delta=0$ (electron-hole pairing does
not occur). Thus, in the paired state the filling factor imbalance
does not depend on temperature  and is given by Eq. (\ref{39}).

Equations (\ref{37}) yield the  mean-field critical temperature of
pairing\cite{32}
\begin{equation}\label{40}
    T_{mf}=\frac{J_{12}}{2}\frac{|\tilde{\nu}|}{\ln
\frac{1+|\tilde{\nu}|}{1-|\tilde{\nu}|}},
\end{equation}
where $\tilde{\nu}$ is given by Eq. (\ref{39}).

The temperature $T_{mf}$ is the function of $d/\ell$. For a given
sample the distance $d$ is fixed and the parameter $d/\ell$ depends
only on the magnetic field. Therefore, it is instructive to present
the critical temperature as the function of the magnetic field. We
choose the $B_d=\phi_0/\pi d^2$ units for $B$, where $\phi_0=h c/2
e$ is the magnetic flux quantum [$B/B_d=(d/\ell)^2$]. The quantity
$e^2/d$ is used as the energy unit. In computation $T_{mf}$ we
account for the dynamical stability condition, implying that
$T_{mf}=0$ at $d/\ell>\tilde{d}_{c1}$. The result of computations
for two types of heterostructures is presented in Fig.\ref{f2}.

\begin{figure}
\begin{center}
\includegraphics[width=8cm]{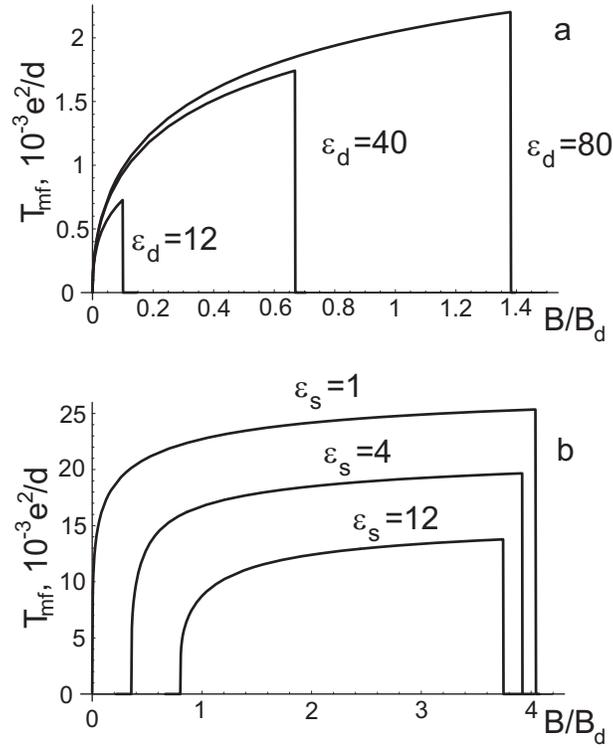}
\end{center}
\caption{The dependence of the mean-field critical temperature on
the magnetic field for the TI - dielectric film - TI heterostructure
(a) and for the heterostructure "TI film on a dielectric
substrate"(b).} \label{f2}
\end{figure}

The critical temperature $T_{mf}$ is not a temperature of the
superfluid transition. The superfluid transition temperature is
given by the Kosterlitz-Thouless equation
\begin{equation}\label{41}
    T_s=\frac{\pi}{2}\rho_s(T_s),
\end{equation}
where $\rho_s(T)$ is the superfluid stiffness. Equation (\ref{41})
can be applied under assumption that the gas of bound electron-hole
pairs exists. The latter requires $T_s< T_{mf}$. Evaluation of $T_s$
shows (see below) that, actually, the strong inequality $T_s\ll
T_{mf}$ is fulfilled. The temperature $T_{mf}$ can therefore be
interpreted as an analog of the ionization temperature. One can see
from Fig. \ref{f2}(b) that the dependence of $T_{mf}$ on $B/B_d$ is
saturated at large $B$. It is connected with that $T_{mf}$ is
determined by the binding energy. The limit $B\gg B_d$ corresponds
to $\ell\ll d$ and in the latter case the binding energy is
determined in the main part by the interlayer distance $d$.

The superfluid stiffness is the coefficient of the expansion of the
free energy in the gradient of the phase of the superfluid order
parameter:
$$F=F_0+\frac{S}{2}\rho_s (\nabla\varphi)^2.$$
We compute $\rho_s(T)$ as follows.  We consider the many-particle
wave function
\begin{equation}\label{42}
    |\Psi\rangle=\prod_X\left(\cos\frac{\theta_X}{2}c^+_{1,X+\frac{Q_y\ell^2}{2}}
+e^{i(Q_x X+\varphi_X)}c^+_{2,X-\frac{Q_y\ell^2}{2}}\right)|0\rangle.
\end{equation}
Equation (\ref{42}) describes the state with a uniform gradient of
the phase of the order parameter : $\nabla\varphi={\bf
Q}=(Q_x,Q_y)$. To see that we neglect for a moment the fluctuations
($\theta_X=\theta_0$ and $\varphi_X=0$) and define the
space-dependent order parameter
\begin{equation}\label{43}
    \Delta({\bf r})=\sum_{X,X'}\psi_X^*({\bf r})\psi_{X'}({\bf
r})\langle\Psi|c^+_{1X}c_{2X'}|\Psi\rangle.
\end{equation}
In (\ref{43})  $\psi_X({\bf
r})=(\sqrt{\pi^{1/2}L_y\ell})^{-1}\exp(-iXy/\ell^2-(x-X)^2/2\ell^2)$
is the one-particle wave function for the zeroth Landau level, and
$L_y$ is the size of the system in the $y$-direction. The direct
calculation yields
\begin{equation}\label{44}
 \Delta({\bf r})=\frac{\sin \theta_0}{2}e^{-\frac{Q^2\ell^2}{2}}e^{i{\bf Q}\cdot{\bf
 r}}.
\end{equation}
Equation (\ref{44}) shows that ${\bf Q}$ is  indeed the gradient of
the phase of the order parameter.

The free energy is given by the formula
\begin{equation}\label{45}
    F=E_0(Q)-T{\cal S}=E_0(Q)+T\sum_{\bf q}\ln\left(1-e^{-\frac{\Omega({\bf q},{\bf
Q})}{T}}\right),
\end{equation}
where
\begin{equation}\label{45a}
 E_0(Q)=E_0(0)+\frac{S}{8\pi\ell^2}\sin^2\theta_0\left[F_D(0)-F_D(Q)\right]
\end{equation}
is the energy of state (\ref{42}),  ${\cal S}$ is the entropy of the
gas of elementary collective excitations, $\Omega({\bf q},{\bf Q})$
is the spectrum of excitations, and $E_0(0)$ is the energy given by
Eq. (\ref{9}).

At ${\bf Q}=0$ the spectrum $\Omega({\bf q},0)=\Omega(q)$ [Eq.
(\ref{25})] is isotropic. Anisotropy of $\Omega({\bf q},{\bf Q})$ is
connected with the appearance of a specific direction in the system
(the direction of the phase gradient).

Considering state (\ref{42}) and repeating the same steps as in
obtaining the spectrum (\ref{25}) we find  $\Omega({\bf q},{\bf Q})$
for ${\bf q}$ directed parallel to the $x$ axis.

The energy of fluctuations has the form (\ref{16}), where the matrix
${\bf K}$ depends on ${\bf Q}$:
\begin{equation}\label{46}
 {\bf K}({\bf q}, {\bf Q})=\left(
 \begin{array}{cc}
 K_{zz}({\bf q},{\bf Q}) & K_{z\varphi}({\bf q},{\bf Q}) \\
 K^*_{z\varphi} ({\bf q},{\bf Q})& K_{\varphi\varphi}({\bf q},{\bf Q}) \\
 \end{array}
 \right)
\end{equation}
with ${\bf q}=q {\bf i}_x$,
\begin{equation}\label{48}
 K_{zz}({\bf q}, {\bf Q})=2\left(H({\bf q},{\bf Q})-
 F_S(q)+F_D(Q)+\cot^2\theta_0\Xi({\bf q},{\bf Q})\right),
\end{equation}
\begin{equation}\label{49}
K_{\varphi\varphi}({\bf q}, {\bf
Q})=\frac{1}{2}\sin^2\theta_0\Xi({\bf q}, {\bf Q}),
\end{equation}
\begin{equation}\label{50}
 K_{z\varphi}({\bf q}, {\bf Q})=i \tilde{K}_{z\varphi}({\bf q}, {\bf Q})=i \cos
\theta_0\left[F_D(|{\bf q}-{\bf Q}|)-F_D(|{\bf q}+{\bf Q}|)\right]/2,
\end{equation}
\begin{equation}\label{51}
    H({\bf q},{\bf Q})=\frac{1}{2\pi\ell^2}\left(\frac{V_{11}(q)+V_{22}(q)}{2}-V_{12}(q)\cos(|{\bf q}\times{\bf Q}|\ell^2)\right)
    e^{-\frac{q^2\ell^2}{2}},
\end{equation}
and
\begin{equation}\label{52}
\Xi({\bf q}, {\bf Q})=F_D(Q)-\frac{F_D(|{\bf q}+{\bf Q}|)+F_D(|{\bf
q}-{\bf Q}|)}{2}.
\end{equation}

The spectrum has the form
\begin{equation}\label{53}
    \Omega({\bf q}, {\bf Q})=\sqrt{K_{zz}({\bf q}, {\bf Q})K_{\varphi\varphi}({\bf q}, {\bf
Q})}+\tilde{K}_{z\varphi}({\bf q}, {\bf Q}).
\end{equation}
One can see that expression (\ref{53}) is invariant with respect to
rotation of the coordinate axes.  The restriction ${\bf q}=q {\bf
i}_x$ can therefore be omitted and Eq. (\ref{53}) yields the
spectrum at the general ${\bf q}$. The spectrum Eq. (\ref{53}) is
anisotropic because of its dependence on the angle between ${\bf q}$
and ${\bf Q}$.

Expanding (\ref{45}) in ${\bf Q}$ we arrive at the following
expression for the superfluid stiffness
\begin{equation}\label{54}
    \rho_s=\rho_{s0}+\delta\rho_s(T),
\end{equation}
where
\begin{equation}\label{55}
    \rho_{s0}=\frac{\sin^2 \theta_0}{8\pi\ell^2} F^{''}_D(Q)|_{Q=0}=\frac{\ell^2}{32\pi^2}\sin^2 \theta_0\int_0^\infty
p^3 V_{12}(p) e^{-\frac{p^2 \ell^2}{2}}d p
\end{equation}
is the zero-temperature superfluid stiffness, and
\begin{equation}\label{56}
    \delta\rho_s(T)=\frac{1}{S}\sum_{\bf q}\left[N_B(q) \frac{\partial^2 \Omega({\bf q},{\bf
Q})}{\partial Q^2}-\frac{1}{T}N_B(q)(1+N_B(q))\left(\frac{\partial
\Omega({\bf q},{\bf Q})}{\partial Q}\right)^2\right]\Bigg|_{Q=0}
\end{equation}
is its temperature correction. In (\ref{56})
$N_B(q)=(e^{\Omega(q)/T}-1)^{-1}$ is the Bose distribution function.

One can show that $\delta\rho_s(T)<0$. We note that Eq. (\ref{56})
generalizes the expression for the superfluid stiffness.\cite{lp}
The case of Ref. \onlinecite{lp} corresponds to a Bose gas in a free
space. In that case the Galilean transformation yields the following
 spectrum of excitations $\Omega({\bf
q})=\Omega_0(q)+\hbar^2 {\bf q}\cdot {\bf Q}/M$ with $\Omega_0(q)$,
the excitation spectrum in the reference frame, where the Bose gas
is at rest, and $M$ is the mass of the Bose particle. Then, Eq.
(\ref{56}) is reduced  to the common one.\cite{lp} The dependence of
the spectrum $\Omega({\bf q},{\bf Q})$ [Eq. (\ref{53})] on ${\bf Q}$
is more complicated and the first term in (\ref{56}) should be taken
into account.

\begin{figure}
\begin{center}
\includegraphics[width=8cm]{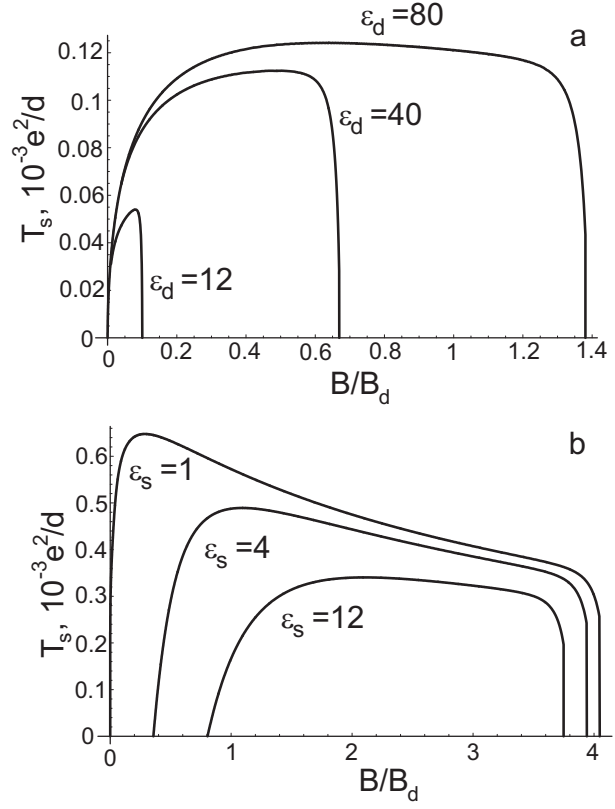}
\end{center}
\caption{The dependence of the superfluid transition temperature on
the magnetic field for the TI - dielectric film - TI heterostructure
(a) and for the heterostructure "TI film on a dielectric
substrate"(b).} \label{f3}
\end{figure}

 The dependence of the superfluid transition temperature on
the magnetic field is shown in Fig. \ref{f3}. One can see that at
all $B$ the inequality $T_s\ll T_{mf}$ is satisfied. The temperature
$T_s$ is a non-monotonic function of the magnetic field. The
superfluid state can be realized in the whole range of allowed
$d/\ell$ ($\tilde{d}_{c1}<d/\ell<\tilde{d}_{c2}$), but at $d/\ell$
near the upper and lower range limit the temperature $T_s$
approaches zero. The maximum temperature is reached in the middle of
that range.

For the TI film structure the temperature of the superfluid
transition is higher if a substrate with a lower dielectric constant
$\varepsilon_s$ is used. For the TI - dielectric film - TI structure
the decrease in the dielectric constant of the film $\varepsilon_d$
results in lowering the superfluid transition temperature.

\section{Conclusion}

In conclusion, we have shown that TI heterostructures are suitable
for realizing the superfluidity of spatially indirect
magnetoexcitons. The structure "TI film on a substrate" is
preferable to the  TI - dielectric film - TI structure. The main
disadvantage of the latter one is that the dielectric layer
separating two TIs should be rather thin. For instance, for $B=1$ T,
$\varepsilon_d=12$ and $\varepsilon_{TI}=80$ the dielectric layer
should not be thicker than 8 nm. For the same parameters the
thickness of the TI film for the substrate with $\varepsilon_s=12$
can be up to 50 nm. The TI film structure has the problem of
shorting two working surfaces through the side surface. But this
problem can be resolved by depositing a magnetic insulator on the
side surface. It opens a gap in the energy spectrum of the side
surface states: $E=\pm\sqrt{m^2+v^2_F p^2}$ (with $m=J_H S_z\gg
T_s$, where $J_H$ is the energy  of the exchange coupling of the
electron and ion spin, and $S_z$ is  the value of the spin of the
magnetic ions) and prevents the interlayer leakage of the
counterflow current through the side surface.

Taking $d=10$ nm we evaluate that the maximum temperature of the
superfluid transition $T_s$ is about 1 K for the TI film structures
and is about 0.2 K for the TI - dielectric film - TI structures.
Graphene heterostructures are characterized by  slightly higher
\cite{18} temperatures of the transition into the superfluid state,
but the disadvantage of graphene heterostructures is that the
magnetoexciton superfluidity can be realized only under application
of the interlayer gate voltage.

\section*{Acknowledgment}
This work was supported by the Ukraine State Program
"Nanotechnologies and nanomaterials" Project No. 1.1.5.21.

\appendix
\section{Landau level eigenfunctions}

In the Landau gauge ${\bf A}=(0,x B,0)$ Eq. (\ref{1}) yields the
following eigenfunctions.

For the zeroth Landau level
\begin{equation}\label{2}
    {\bm \Phi}_{0,X}({\bf r})=\frac{e^{-i k y}e^{-\frac{(x-X)^2}{2
    \ell^2}}}{\pi^{1/4}\sqrt{L_y \ell}}\left(
                                         \begin{array}{c}
                                           0 \\
                                           1 \\
                                         \end{array}
                                       \right),
\end{equation}
where $k$ is the wave number connected with the guiding center of
the orbit $X$ by the relation $k=X/\ell^2$.

For nonzero levels
\begin{eqnarray}\label{3}
 {\bm \Phi}_{\pm N,X}({\bf r})=\frac{e^{-i k y}e^{-\frac{(x-X)^2}{2
    \ell^2}}}{\pi^{1/4}\sqrt{2^{N}N!L_y \ell}}\frac{1}{\sqrt{\frac{(E_{\pm N}-m)^2}{2 N}
    +\left(\frac{\hbar v_F}{\ell}\right)^2}}\left(
             \begin{array}{c}
          (E_{\pm N}-m) H_{N-1}(\frac{x-X}{l}) \\
             \pm \frac{\hbar v_F}{\ell}H_{N}(\frac{x-X}{l}) \\
                     \end{array}
           \right),
\end{eqnarray}
where $H_N(x)$ are the Hermite polynomials.

In a moderate magnetic field and in the absence of magnetic exchange
interactions the strong inequality $m\ll \hbar v_F/\ell$ is
fulfilled and the relation between the components of the
eigenfunction  (\ref{3}) is practically the same as for the Landau
level eigenfunctions in graphene.

The important difference between graphene and a topological
insulator is that the zeroth Landau level on the TI surface is
completely spin polarized. Due to the square root dependence of the
Landau level energies on the magnetic field even in a rather small
field the states with an admixture of the opposite spin polarization
(e.g., the states in the $N=1$ Landau level) are separated from the
zero level spin-polarized states by a large energy gap.

\end{document}